\documentclass[aps,prl,twocolumn,superscriptaddress,preprintnumbers,nobibnotes]{revtex4-1}

\usepackage{graphicx}
\usepackage{amssymb,amsfonts,amsmath}
\usepackage{hyperref}

\newcommand{\cut}[1]{\null}
\newcommand{\add}[1]{#1}

\renewcommand{\emph}[1]{{\it#1}}
\newcommand{\fig}[1]{\textbf{Fig.~\ref{#1}}}
\newcommand{\movie}[1]{\textbf{Movie~#1}}
\newcommand{\eq}[1]{\textbf{Eq.~\ref{#1}}}
\newcommand{\sect}[1]{\textbf{\S~\ref{#1}}}

\renewcommand{\vec}[1]{\boldsymbol{#1}}
\newcommand{\tens}[1]{\boldsymbol{#1}}
\newcommand{\bnabla}{\vec{\nabla}}
\newcommand{\phalf}{$+1/2$ }
\newcommand{\mhalf}{$-1/2$ }
\newcommand{\pmhalf}{$\pm1/2$ }
\newcommand{\av}[1]{\left\langle #1 \right\rangle}
\DeclareSymbolFont{newfont}{OML}{cmm}{m}{it}
\DeclareMathSymbol{\Epsilon}{3}{newfont}{15}

\begin{document}

\preprint{to be published in \textit{Soft Matter}}

\title{Dancing disclinations in confined active nematics}


\author{Tyler N. Shendruk}
\thanks{These authors contributed equally to this work.}
\affiliation{The Rudolf Peierls Centre for Theoretical Physics, 1 Keble Road, Oxford, OX1 3NP, United Kingdom}
\affiliation{Center for Studies in Physics and Biology, The Rockefeller University, 1230 York Avenue, New York,  10065, USA}

\author{Amin Doostmohammadi}
\thanks{These authors contributed equally to this work.}
\affiliation{The Rudolf Peierls Centre for Theoretical Physics, 1 Keble Road, Oxford, OX1 3NP, United Kingdom}

\author{Kristian Thijssen}
\thanks{These authors contributed equally to this work.}
\affiliation{Department of Applied Physics, Eindhoven University of Technology, 5600 MB, Eindhoven, Netherlands}

\author{Julia M. Yeomans}
\email[Corresponding author: ]{julia.yeomans@physics.ox.ac.uk}
\affiliation{The Rudolf Peierls Centre for Theoretical Physics, 1 Keble Road, Oxford, OX1 3NP, United Kingdom}

\date{\today}

\begin{abstract}
The spontaneous emergence of collective flows is a generic property of active fluids and often leads to chaotic flow patterns characterised by swirls, jets, and topological disclinations in their orientation field. 
However, the ability to achieve structured flows and ordered disclinations is of particular importance in the design and control of active systems. 
By confining an active nematic fluid within a channel, we find a regular motion of disclinations, in conjunction with a well defined and dynamic vortex lattice. 
As pairs of moving disclinations travel through the channel, they continually exchange partners producing a dynamic ordered state, reminiscent of Ceilidh dancing. 
We anticipate that this biomimetic ability to self-assemble organised topological disclinations and dynamically structured flow fields in engineered geometries will pave the road towards establishing new active topological microfluidic devices.
\end{abstract}

\pacs{}

\maketitle

\section{Introduction}


Bacterial suspensions~\cite{Wensink2012}, cellular monolayers~\cite{Benoit2012} and sub-cellular filament/motor-protein mixtures~\cite{Chate2012,Dogic2012} are continuously driven far from equilibrium through intrinsic energy injection by their biological constituent elements. 
This microscopic energy input can result in the spontaneous emergence of large-scale collective behaviour, including flocking~\cite{TonerTu,Ramaswamy2015,Attanasi2015}, unidirectional flows~\cite{Joanny2005}, meso-scale turbulence~\cite{Dombrowski2004,Wensink2012,creppy2015}, topological defect pair production~\cite{Giomi2013,Thampi2014epl,Giomi2014}, and phase separation~\cite{Baskaran2014}. Moreover there is a rapidly growing list of biological systems, including anterior-posterior establishment by active cytoplasmic streaming in oocytes~\cite{Trong2015}, pupal wing morphogenesis~\cite{Etournay2015}, wound healing~\cite{Silberzan2007,Wolgemuth2011}, and cancer invasion~\cite{Wolf2007,Tse2012}, that has been identified as leveraging collective flows for active self-organisation. Recent works have shown that nematic disclinations in growing bacterial colonies can determine morphological changes~\cite{Doostmohammadi2016PRL} and play a crucial role in the dynamic structure of cell groups such as in ameoboid cells~\cite{Gruler2000,Silberzan2014} and spindle-shaped cells~\cite{Silberzan2016}, and the emergence of 3D mounds in layers of stem cells~\cite{Sano2016}. 

These examples suggest the possibility of harnessing active driving by manufacturing active microfluidic devices. In a few engineered instances, spontaneous collective motion has been controlled through geometrical constraints. 
Ratchet systems have been constructed to sort cells against entropy~\cite{Austin2007,Shear2009,Mahmud2009}, while individual asymmetric gears~\cite{Hiratsuka2006,Leonardo2009,Leonardo2010,Sokolov2010} and arrays of smooth rotors~\cite{Thampi2016} have been designed to have persistent rotation when submersed in bacterial baths. 
Long-range static orientational ordering of nematic disclinations has recently been probed in experiments on filament/motor protein mixtures~\cite{Dogic2015} and theoretical works have predicted positional disclination ordering in crystal-like configurations due to hydrodynamic screening in over-damped active nematics~\cite{Doostmohammadi2016NatComm,Putzig2016,Oza2016,Marchetti2016}. 
Nevertheless, the ability to control disclination dynamics in active matter is currently lacking and is a necessary prerequisite to the control of \emph{active topological microfluidics}. 

\begin{figure*}
	\centering
	\includegraphics[width=0.95\textwidth]{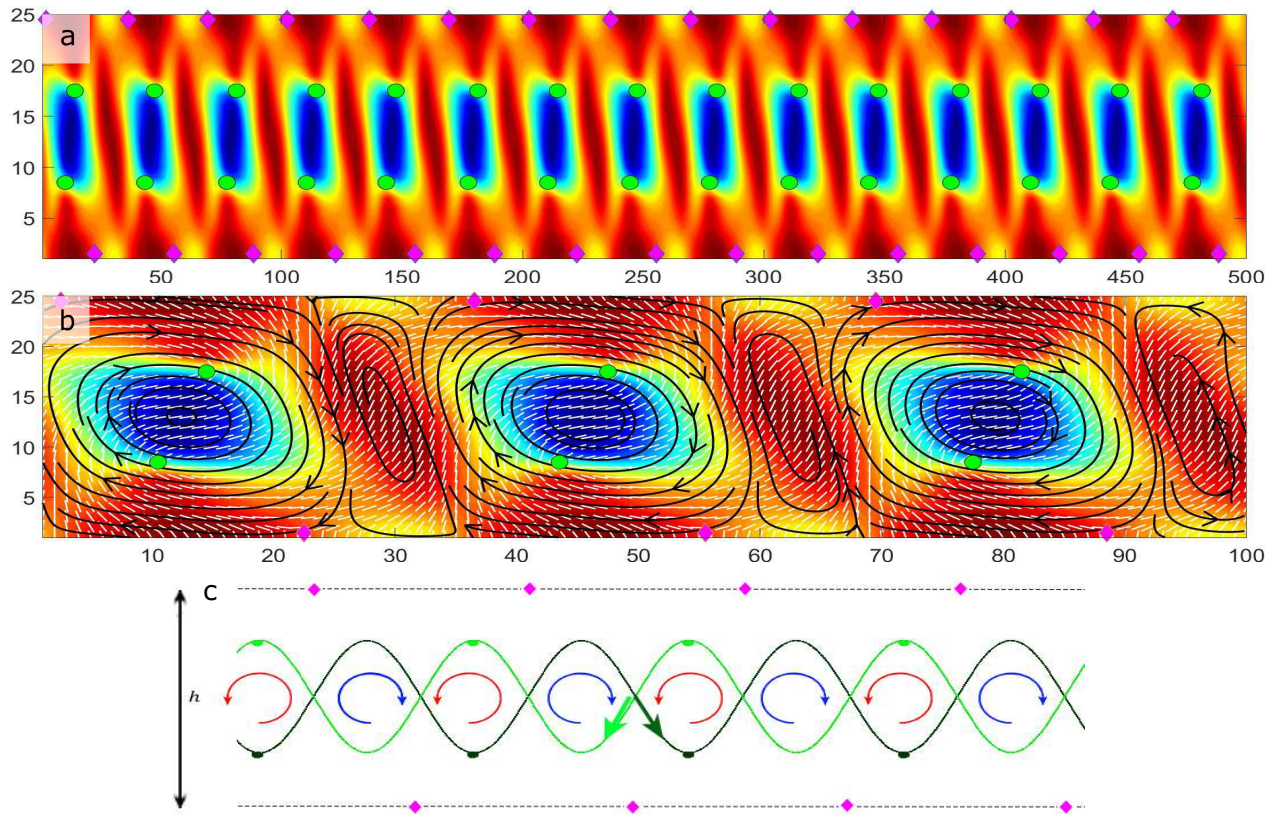}
	\caption{{\bf Dancing disclinations: Topological dynamics (see \movie{1}).} (a) Simulation snapshot of the instantaneous vorticity field, which forms a highly ordered vortex lattice and a dynamically ordered state of disclinations within the channel. Negative ($-1/2$) topological disclinations settle in the vicinity of the channel walls (magenta diamonds), while positive ($+1/2$) disclinations (green dots) transverse the mid-region as they move along the vortex lattice. 
	(b) Zoom showing streamlines and director field. 
	(c) Schematic of ideal Ceilidh dynamics. In a channel of height $h$, \mhalf topological disclinations (magenta diamonds) reside in the near-wall regions where they oscillate weakly about their average position. Within the mid-channel region is the vortex lattice of counter-rotating neighbouring vortices. 
	Moving along the edge of the vortices are the \phalf disclinations (green dots). 
	The system has zero net-topological charge and ideally equal numbers of \phalf disclinations travel to the left as to the right (light and dark green, respectively).}
	\label{fig:dance}
\end{figure*}

Many dense active systems are well modelled as active nematics, and confining such an active nematic is known to result in spontaneous symmetry breaking leading to self-sustained, unidirectional laminar flows at moderate activities~\cite{Joanny2005,Goldstein2012,Miha2013,Edwards2009}. 
It is further known that such unidirectional active flows transition to oscillatory flows at sufficiently high activity~\cite{Giomi2012,Julicher2016} and to meso-scale turbulence at higher activities yet~\cite{Wensink2012,Dunkel2013,Thampi2014epl,Thampi2014,Giomi2015}. 
Linear stability analysis has been performed on the formation of oscillations in the laminar flow with moderate activity~\cite{Giomi2012} but increased activity causes non-linear effects to quickly dominate. 
It has recently been shown~\cite{Doostmohammadi2016turb} that moderate activities in confining (quasi-1D) geometries can produce a system-spanning lattice of flow vortices intermediate to oscillating flows and meso-scale turbulence. 
The vortex lattice has been observed in simulations of sheared active fluids~\cite{slomka2016}, as well as experimentally in quasi-1D suspensions of bacteria~\cite{Wioland2016} and associated simulations~\cite{Wioland2016,Theillard2016}. 
\add{Similarly, spatially ordered textures and disclination lattices have previously been noted in confined and flowing passive liquid crystals~\cite{Grecov2003,Gupta2005,Rey2010}.} 

Here, we confine an active nematic within a two dimensional micro-channel. 
The results of our simulations show that the intermediate activity state is characterized by \emph{dancing disclination dynamics}, where the stationary vortex lattice is accompanied by a dynamically ordered state of disclinations (\fig{fig:dance}(a); \movie{1}). 
The disclination dynamics accompanying the vortex lattice have not hitherto been reported. 
These dancing disclinations are positively charged, long lived, and continually navigate through the channel. 
As they move past each other on the vortex lattice, the positive disclinations form short-lived pairs, leading us to dub this state \emph{topological Ceilidh dynamics}, in rough analogy to the traditional Gaelic dance in which participants form two inward-facing parallel lines and pairs of dancers continually exchange partners as they circulate between the lines. 
Our results for quasi-1D active nematics show that, in this simple geometry, ordered dancing-disclination dynamics emerge as a system-spanning intermediate regime between unidirectional flow and meso-scale turbulence. 

We next treat the Ceilidh dynamics state as a spatially ordered configuration between the unidirectional flow state and meso-scale turbulence. 
As a system-spanning ordered state, the Ceilidh dynamics pattern can itself possess irregularities, which we term \emph{lattice defects}. 
A subset of these lattice defects are found to drive a well-defined net flux that increases incrementally with the quantized number of \emph{drift-lattice defects}.
The dynamically ordered topological and steady-flow structures of Ceilidh dynamics within a channel represent an ideal system for studying the emergence of ordered dynamical structures in active matter. 

\section{Methods}\label{sec:methods}
Many living fluids can be modelled as continuous active nematic liquid crystals because they generally consist of a dense suspension of many (continuum) shape-anisotropic (nematic) self-propelled (active) particles moving through a fluid medium (force-free)~\cite{Marchetti2013}. 
We utilize numerical simulations of active nematohydrodynamics to solve for the density, velocity (with its associated vorticity $\vec{\omega}\left(\vec{r},t\right)$), and orientation tensor $\tens{Q}(\vec{r},t)$ (with associated director $\vec{n}\left(\vec{r},t\right)$ and scalar order $q\left(\vec{r},t\right)$) fields~\cite{Marenduzzo2007}. 
We assume that the nematic fluid has a single nematic elastic constant $K$ and that the activity coefficient $\zeta>0$, representing extensile active systems such as pusher-type bacteria~\cite{Lauga2009}. 
The 2D active nematic is confined between two parallel no-slip channel walls separated by a distance $h$, producing a quasi-1D system. 
Strong anchoring at the boundaries sets the director \cut{parallel}\add{perpendicular} to the confining walls. 

\subsection{Governing equations}\label{sec:theory}
Active nematohydrodynamics have been extensively applied to intrinsically out-of-equilibrium biological systems composed of rod-like constituents, including bacterial suspensions~\cite{Volfson2008}, filament/motor protein mixtures~\cite{Dogic2012,Giomi2013,Thampi2013}, and cellular monolayers~\cite{Julicher2008,Silberzan2014,Doostmohammadi2015}. 
The continuum fields that must be solved are the total density $\rho$, the velocity $\vec{u}$, and the orientation $\tens{Q}$ fields. 
The orientational order of the fluid is described by the nematic tensor $\tens{Q}=\frac{3q}{2}( \vec{n}\vec{n}-\tens{I}/3)$, where $q$ denotes the magnitude of the orientational order, $\vec{n}$ is the director, and $\tens{I}$ the identity tensor~\cite{DeGennesBook}. 

The transport equation for the nematic tensor field is~\cite{Berisbook}
\begin{align}
\left(\partial_t + \vec{u}\cdot\bnabla\right) \tens{Q} - \tens{S} &= \Gamma \tens{H},
\label{eqn:lc}
\end{align}
where $\Gamma$ is a rotational diffusivity and $\tens{S}$ is the co-rotational advection term that accounts for the impact of the strain rate $\tens{E}=(\bnabla\vec{u}^{T}+\bnabla\vec{u})/2$ and vorticity $\tens{\Omega}=(\bnabla\vec{u}^{T}-\bnabla\vec{u})/2$ on the director field.
The co-rotational advection has the form
\begin{align}
\tens{S} &= \left(\lambda \tens{E} + \tens{\Omega}\right)\cdot\tens{\mathcal{Q}} + \tens{\mathcal{Q}}\cdot\left(\lambda \tens{E} - \tens{\Omega}\right) - 2 \lambda \tens{\mathcal{Q}}\left( \tens{Q} : \bnabla \vec{u}\right), 
 \label{eqn:cor}
\end{align}
where $\tens{\mathcal{Q}} \equiv \tens{Q} + \tens{I} / 3$ and the alignment parameter $\lambda$ controls the degree of coupling between the orientation field and velocity gradients and determines the objective time derivative of orientation, with $\lambda=\pm1$ corresponding to an upper and lower convected derivative respectively. 
The alignment parameter determines whether the nematogens align or tumble in a shear flow\cite{Thampi2014}. 
The relaxation of the orientational order is controlled by the free energy $\mathcal{F}$\add{$=\int f dV$} through the molecular field,
\begin{align}
\tens{H} &= -(\frac{\delta f}{\delta\tens{Q}} - \frac{1}{3}\tens{I} \; \text{Tr}\frac{\delta f}{\delta\tens{Q}}). \label{eqn:molpot}
\end{align}
The free energy \add{density} has two components, which are the Landau-de Gennes bulk free energy \add{density} and the elastic free energy \add{density} due to spatial inhomogeneities in the order parameter \cite{DeGennesBook}:
\add{\begin{align}
f &=  \frac{\mathcal{A}}{2} Q^2 + \frac{\mathcal{B}}{3} Q^3 + \frac{\mathcal{C}}{4} Q^4 + \frac{K}{2}\left(\nabla Q\right)^2, 
\end{align}}
assuming a single elastic Frank coefficient $K$. 

The local density and velocity field obey the incompressible Navier-Stokes equations
\begin{align}
\bnabla\cdot\vec{u} &=0,\label{eqn:cont}\\
\left(\partial_t + \vec{u}\cdot\bnabla\right) \vec{u} &= \bnabla\cdot\tens{\Pi} / \rho,
\label{eqn:NS}
\end{align}
where $\tens{\Pi}$ is the generalized stress tensor that includes both nematic and active contributions, in addition to the viscous stress $\tens{\Pi}^\textmd{visc} = 2 \eta \tens{E}$. 
The stress due to elastic contributions arising from nematic ordering within the liquid crystal is
\begin{align}
\tens{\Pi^{\text{elastic}}} &= -P\tens{I} +2 \lambda \tens{\mathcal{Q}} (\tens{Q}:\tens{H}) -\lambda \tens{H}\cdot\tens{\mathcal{Q}}  - \lambda \tens{\mathcal{Q}} \cdot \tens{H} \nonumber\\
                            &\quad -\tens{\nabla}\tens{Q} : \frac{\delta f}{\delta \tens{\nabla}\tens{Q}} + \tens{Q}\cdot\tens{H} - \tens{H}\cdot\tens{Q},
\label{eqn:elastic}
\end{align}
which includes the isotropic pressure $P$~\cite{Berisbook}. 
The active stress accounts for changes in the flow field caused by continual energy injection at the microscopic scale. 
Activity generates flows for nonzero gradients of $\tens{Q}$ and takes the form~\cite{Simha2002}
\begin{align}
\tens{\Pi^{\text{act}}} &= -\zeta \tens{Q}.
\label{eqn:active}
\end{align}
The activity parameter $\zeta$ determines the strength of the active flows with positive and negative values denoting extensile and contractile fluids, respectively. 
\subsection{Numerical implementation}\label{sec:numerics}
The equations of active nematohydrodynamics (\eq{eqn:lc}-\ref{eqn:cont}) are solved using a hybrid lattice Boltzmann and finite difference method~\cite{Marenduzzo2007,Fielding2011,Thampi2014}, with the discrete space and time steps defining the \cut{simulations}\add{simulation} units. 
\add{Parameters can be mapped to physical units that correspond to microtubule-kinesin bundles~\cite{Dogic2012}, dense bacterial suspensions~\cite{Wensink2012}, or other active nematics of interest when the material properties are well known~\cite{Cates2008,Thampi2013}.}
Simulations presented here use the parameters $\Gamma=0.34$, $K=0.04$, $\eta=2/3$, and $\rho=1$, except where explicitly stated otherwise. 
\cut{These parameters are chosen to match those used to study \cut{microtuble}\add{microtubule} bundles~\cite{Dogic2012}.} 
The alignment parameter is taken to be $\lambda=0.3$, which is in the flow tumbling regime. 
The active fluid is extensile with $\zeta>0$, though Ceilidh dynamics were also observed for $\zeta<0$ and $\lambda<0$. 
Unless otherwise stated, $\zeta=0.04$. 
The Landau-de Gennes bulk free energy parameters are chosen to be $\mathcal{A} = 0$, \add{$\mathcal{B} = -0.3$} and $\mathcal{C} = −0.3$ in lattice Boltzmann units. 
The bulk energy density scale is therefore identified \add{from these choices~\cite{Marenduzzo2007}} to be \cut{$\mathcal{E}_G\simeq\mathcal{B}=\mathcal{C}=0.3$} \add{$\mathcal{E}_G\simeq0.1$}. 

\begin{figure}
	\centering
	\includegraphics[width=0.475\textwidth]{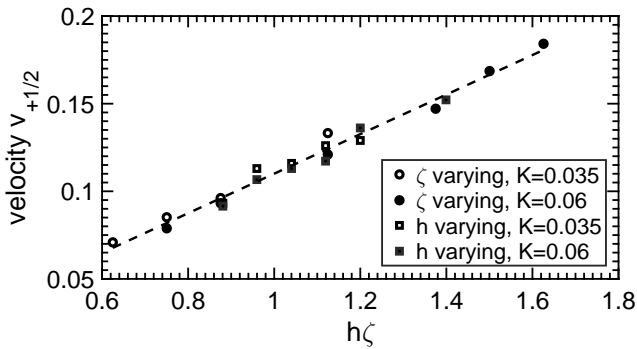} 
	\caption{{\bf Average \phalf disclination speeds.} The average speed $v_{+1/2}$ increases linearly with activity and with height, $v_{+1/2} \sim h\zeta$. The dashed green line shows the scaling predicted by Giomi \textit{et al.}~\cite{Giomi2015}.}
	\label{fig:speed}
\end{figure}

All simulations are \cut{done}\add{performed} in a 2D channel of height $h=25$, unless otherwise stated. 
The walls enforce a no-slip condition on the velocity field through bounce-back boundary conditions and strong \cut{planar}\add{homeotropic} anchoring of the orientation field\cut{, unless otherwise stated}. 
The channel runs in the $\hat{x}$-direction with periodic boundary conditions. 
The channel length is $L=250$ (\fig{fig:speed}-\textbf{\ref{fig:vortexFit}} and \movie{1}-\textbf{2}), $L=300$ (\movie{3}), $L=750$ (\movie{4}), or $L=1000$ (\fig{fig:enstrophy}-\textbf{\ref{fig:driftdefect}}). 
\add{Multiple channel lengths ($L=\left\{100,250,300,500,750,1000\right\}$) are utilized in \fig{fig:driftdefect}(a).}  
\add{The fluid is initialized with a zero velocity and with the director everywhere perpendicular to the walls. 
Initial simulations showed steady state dynamics are reached in $\sim10000$ time steps so an initial warmup of $50000$ is utilized throughout.
}

\section{Results}
For activities where a lattice of flow vortices is stable, the disclinations exist as permanent pairs of oppositely charged \pmhalf topological defects~\cite{DeGennesBook} (\fig{fig:dance}(a); \movie{1}). 
Each \mhalf disclination lingers at a point in the vicinity of the channel wall, while positive \phalf disclinations continually dance; navigating through the channel and following approximately sinusoidal trajectories (\fig{fig:dance}(b)). 
Half of the positive disclinations travel in one direction along the channel and the other half travel along the mirrored trajectory, in the ideal case (\fig{fig:dance}(c)). 
As they move past each other on the vortex lattice, the positive disclinations form ephemeral pairs of dancers that continually exchange partners and periodically skew the vortices as they circulate between the \mhalf disclinations (\fig{fig:dance}(b); \movie{1}).
\subsubsection{Mid-channel vortex lattice and \phalf disclinations}\label{sec:midchannel}
Plus half disclinations in extensile active nematics self-propel themselves along their comet-like tail\cut{.}~\cite{Giomi2014} and move along the borders of counter-rotating vortices (\fig{fig:dance}(b); \movie{1}). 
Although elastic interactions with the near-wall \mhalf disclinations attract the \phalf disclinations towards the bounding planes, the vorticity field and the motility of the \phalf disclinations keep them from leaving the mid-region. 
A disclination cannot switch paths, since this would require overcoming both the deformation free energy barrier of crossing the region between the two paths and also of reorientation. 

\begin{figure}
	\centering
	\includegraphics[width=0.475\textwidth]{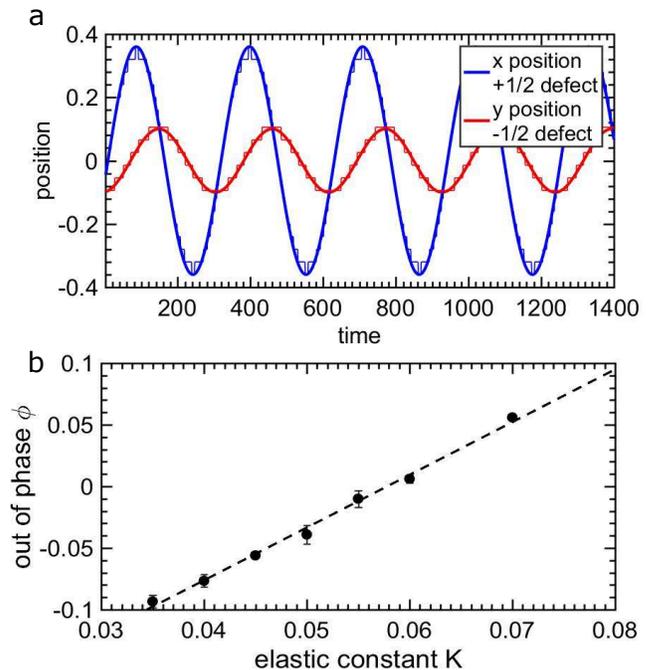}
	\caption{{\bf Trajectories of \pmhalf disclinations about their central position normalized by their amplitude as a function of time in channels of height $h=35$.} 
	(a) Red line denotes lateral \phalf disclination trajectory about the centre of the channel and blue line denotes axial component of the \mhalf disclination trajectory about the mean position. 
	Simulation data (thin lines) are well described as sinusoidal (thick lines). 
	(b) Phase difference between \phalf and \mhalf trajectories as a function of nematic elasticity $K$. 
	}
	\label{fig:phase}
\end{figure}

Increasing the active energy input in the system, increases the speed of the \phalf disclinations, which is seen to increase with the activity as $v_{+1/2}\sim\zeta$ (\fig{fig:speed}). 
The no-slip boundary conditions viscously dissipate this energy and we see that increasing the wall separation $h$ also increases the disclination velocity linearly (\fig{fig:speed}). 
These dependencies are in agreement with the theoretically predicted characteristic self-motility of \phalf disclinations~\cite{Giomi2014} of $v_{+1/2} \sim h \zeta /\eta$ where the channel height $h$ is taken to be the characteristic system size. 
This agreement indicates that nematic elasticity does not dominate the dynamics. 
In fact, increasing the Frank coefficient from $K=0.035$ to $0.07$ produces no observable effect (\fig{fig:speed}). 
For $K$ outside of this region, we do not see topological Ceilidh dynamics for the simulated channel height and activity. 
\add{For $K<0.035$ values we instead observe meso-scale turbulence, while for $K>0.07$ we find oscillating flows --- the dynamical transitions to and from Ceilidh dynamics, and the phase diagram of dynamical steady flow regimes will be discussed in more detail in \sect{sec:transition} and \sect{sec:phaseDiagram}.}

\subsubsection{Near-wall region and \mhalf disclinations}
\label{sec:nearwall}
Unlike \phalf disclinations, \mhalf disclinations in active nematics are not self-motile~\cite{Giomi2014} and their dynamics are primarily set by elastic and hydrodynamic interactions. 
Line tension pins the \mhalf disclinations to the confining walls~\cite{Denniston1996}; however, the \mhalf disclinations can never reside exactly on the surfaces because the director field is imposed by strong homeotropic anchoring boundary conditions, which ensure that the field cannot vary substantially in the near-wall region. It is noteworthy that the dancing dynamics in the mid-channel is independent of the anchoring conditions on the walls. 

Although the \mhalf disclinations do not transverse the channel, they move in-plane slightly about their pinning point and  in a way that is directly correlated to the oscillatory trajectories of the \phalf disclinations (\movie{1}). 
The in-plane oscillations of the \mhalf disclinations parallel to the channel walls have the same constant period as the lateral component of the \phalf disclinations (\fig{fig:phase}(a)), although the amplitude is much smaller. 
Interestingly, the oscillations of the non-motile \mhalf disclinations are generally slightly out-of-phase with the self-propelled \phalf disclinations. 
This phase difference results from the nematic elasticity $K$ (\fig{fig:phase}(b)) but the fact that it is small again suggests that, although present, nematic elasticity is not a significant factor in these simulations of topological Ceilidh dynamics. 

\begin{figure}
	\centering
	\includegraphics[width=0.475\textwidth]{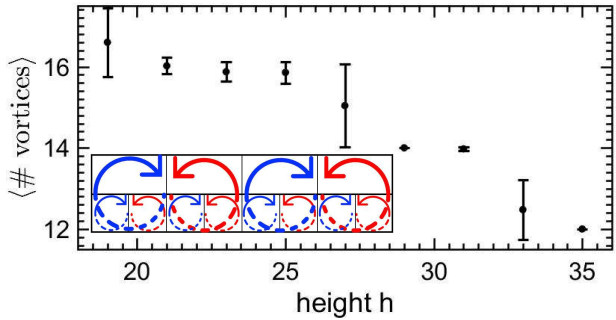}
	\caption{{\bf The vortex lattice fills the channel.} Increasing channel height decreases the number of vortices per unit length. 
	The number of vortices decreases stepwise in increments of two since counter-rotating vortices must occur in pairs. 
	Non-integer intermediate values represent averages of the even numbers of vortices and the error bars represent the variance. 
	(inset) Schematic representation showing the number of vortices per unit length decrease as the channel height is increased since pairs of counter-rotating vortices occupy a larger area.}
	\label{fig:vortexFit}
\end{figure}


\subsection{Dynamical transitions to and from Ceilidh dynamics}
\label{sec:transition}

We treat the flows (unidirectional, oscillating, Ceilidh dynamics and meso-scale turbulence) as ordered states that arise from a series of dynamical transitions as a function of dimensionless activity (\movie{2}). 
The marked difference between laminar flows (\movie{2}(a)) and the intermediate Ceilidh dance state (\movie{2}(c)) is the formation and structure of the vortex lattice. 
As the activity is increased, we first observe oscillating but unidirectional laminar flow, with temporary traveling vorticity patterns that intermittently dissipate then re-form (\movie{2}(a)). 
For the vortex lattice to form, the size of the vortices must be commensurate with the channel height (\fig{fig:vortexFit}(inset)). 
When the height of the channel is increased then the number of vortices per unit length decreases (\fig{fig:vortexFit}). 
Because vortices must occur in counter-rotating pairs, we find that increasing the channel height (for constant channel length) eventually results in a decrease of two in the number of vortices (\fig{fig:vortexFit}). 

\begin{figure}
	\centering
	\includegraphics[width=0.475\textwidth]{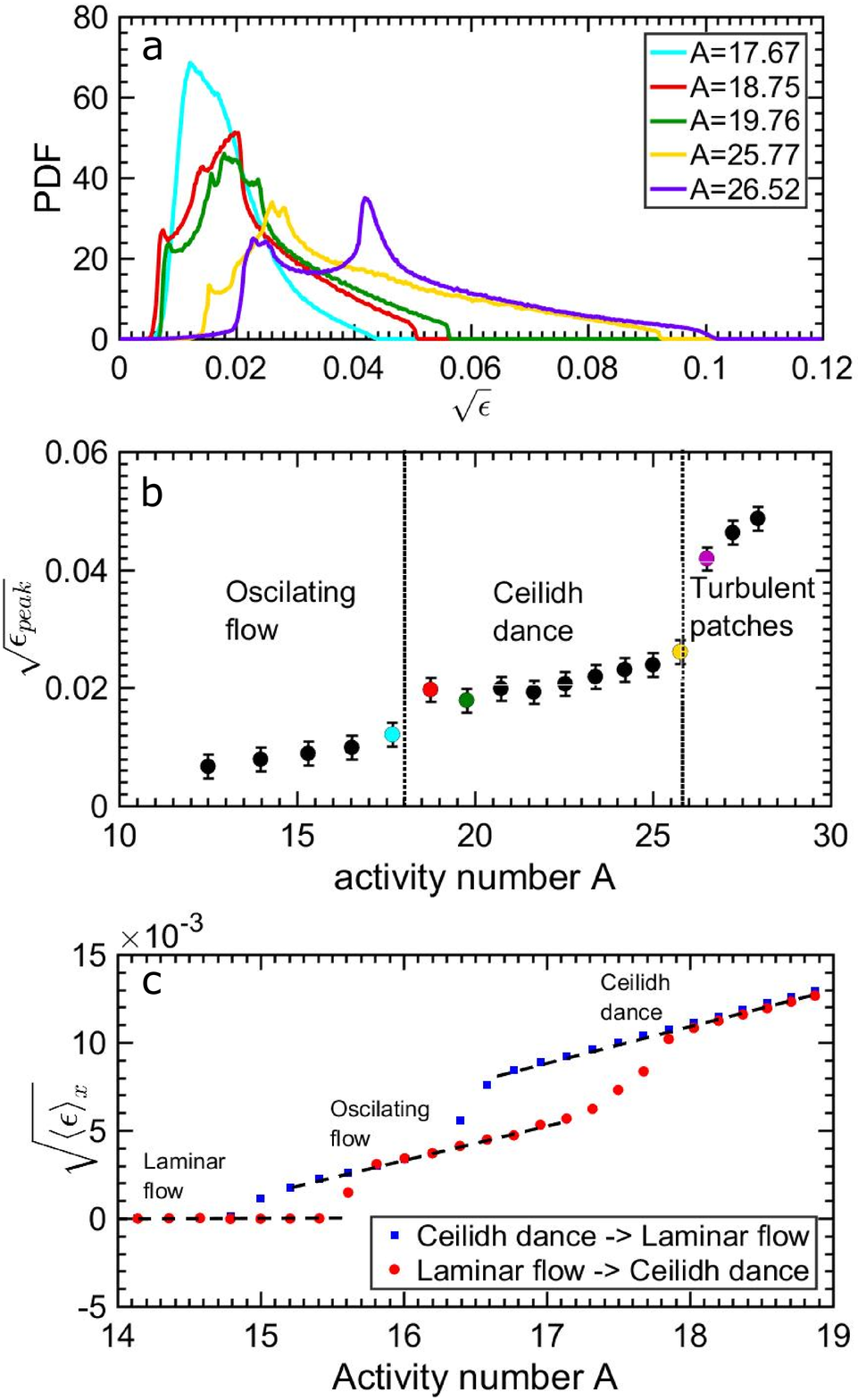} 
	\caption{{\bf Enstrophy} (a) \cut{Channel}\add{Square root of the channel height}-averaged enstrophy \add{$\Epsilon = \av{\vec{\omega}\left(\vec{r},t\right) \cdot \vec{\omega}\left(\vec{r},t\right)}_y$} distribution for different values of the dimensionless activity number $A = \sqrt{\zeta h^2/K}$. 
	(b) \cut{Enstrophy}\add{Channel height-averaged enstrophy distribution} peak position $\Epsilon_\text{peak}$ as a function of $A$.
	Coloured markers correspond to the distributions shown in (a).
	(c) Activity is increased in incremental steps of $\Delta\zeta=2\times10^{-4}$, from non-flowing to oscillating flow to Ceilidh dynamics. The linear rise \add{of the system-averaged enstrophy $\av{\Epsilon}_x$} within the unidirectional laminar flow state is subtracted off and the dynamical transitions to Ceilidh dynamics are seen to have measurable hysteresis. Red symbols denote increasing activity, whereas blue symbols denote decreasing activity. 
	}
	\label{fig:enstrophy}
\end{figure}

We characterize the transition between the oscillating state and the Ceilidh dance (vortex lattice) regimes via the enstrophy\cut{, and use the first moment of the enstrophy distribution $\Epsilon$ as the order parameter for the Ceilidh dance. 
The enstrophy averaged across the channel represents the instantaneous strength of the vorticity at every point along the channel, $\Epsilon\left(x,t\right) = \left\langle \vec{\omega}\left(\vec{r},t\right) \cdot \vec{\omega}\left(\vec{r},t\right) \right\rangle_y$.\P We }\add{ $\vec{\omega}\left(\vec{r},t\right) \cdot \vec{\omega}\left(\vec{r},t\right)$, and } 
measure activity in terms of the dimensionless activity number $A = \sqrt{\zeta h^2/K}$, which is motivated in more detail in the next section.
At low activities, (in the unidirectional flow state; \movie{2}(a)), the distribution of the \cut{channel-averaged} enstrophy \add{averaged across the channel $\Epsilon\left(x\right) = \av{\vec{\omega}\left(\vec{r},t\right) \cdot \vec{\omega}\left(\vec{r},t\right)}_y$} is sharply peaked around small values (not shown). 
At moderate activities, the system enters the oscillating-flow state and the \add{channel height-averaged} enstrophy distribution spreads to higher values (\fig{fig:enstrophy}(a); blue line). 
In this oscillating-flow state, the distribution is seen to decay monotonically from a maximum likelihood at its smallest non-zero value. 
At higher activity, however, the peak \add{value $\Epsilon_\text{peak}$} shifts from the distributions smallest non-zero value to larger values (\fig{fig:enstrophy}(a); red line). 
This peak-shift is due to the appearance of the vortex lattice and is a good indicator of the state of the flow structure (\fig{fig:enstrophy}(b)). 

A further sharp rise in the position of the enstrophy maximum in Fig.~{\ref{fig:enstrophy}(b) marks the onset to the transition between the Ceilidh dynamics and meso-scale turbulence (\movie{2}(d)). 
At this point the first signals of localized turbulent patches are identified with short mean lifetimes. 
Simulations of the following transition to system-spanning, steady-state meso-scale turbulence show that the transition belongs to the directed percolation universality class~\cite{Doostmohammadi2016turb}. 
Experiments in microfluidic circuits suggest that the transition may be accompanied by a preferential direction depending on bacteria orientation at the walls~\cite{Wioland2016}. 

To further investigate  the nature of the dynamical transitions between steady states, we consider the transformation from laminar flow to Ceilidh dynamics as a function of activity number \add{by measuring the system-averaged enstrophy $\av{\Epsilon}_x$}. 
As the activity is increased incrementally from zero, the system moves slowly through the dynamical steady states.
When the activity is then lowered from the vortex-lattice state the Ceilidh dance state remains stable to lower activities, indicating hysteresis in this dynamic and intrinsically far-from-equilibrium flowing system (\fig{fig:enstrophy}(c)). 

\subsection{Dynamical phase diagram}
\label{sec:phaseDiagram}

By mapping out the parameter space for which topological Ceilidh dynamics arise, we construct a phase diagram of dynamical steady flow regimes (\fig{fig:phaseDiagram}). 
The competition between the activity driving a flow field and nematic elasticity resisting the associated deformation leads to the characteristic activity-induced length scale $\ell_\zeta \sim \sqrt{K/\zeta}$, which corresponds to the length scale over which the active stresses are accommodated by orientational elasticity~\cite{Ramaswamy2010,Thampi2013}. 
However, confinement truncates the allowable range of $\ell_\zeta$ and screens interactions on separations $>h$. 
Thus, there is a competition between the activity-induced length scale $\ell_\zeta$ and the screening length $h$, as described by the dimensionless activity number $A = h/\ell_\zeta = \sqrt{\zeta h^2/K}$. 
\add{For suspensions of microtubule-kinesin bundles with $\ell_\zeta \sim 100\mu\text{m}$~\cite{Dogic2012} confined within channels of height $100\mu\text{m} \lesssim h \lesssim 1\text{cm}$~\cite{Dogic2015}, we expect a correspondingly broad range of activity numbers $A \sim 1-10^{2}$.}

However, the activity number is not the only control parameter that affects the order parameter \cut{$\av{\Epsilon}$}\add{$\av{\Epsilon}_x$} since the \phalf disclination speed increases linearly with both activity and channel height (\fig{fig:speed}) and, therefore, does not depend directly on $A$. 
In addition to the dimensionless activity number $A$, we must account for the self-motility speed of the \phalf disclinations~\cite{Giomi2014} $v_{+1/2}\sim h\zeta/\eta$. 
The magnitude of the self motility must be judged against the characteristic velocity scale of the nematic liquid crystal~\cite{Hemingway2016} $\nu_Q=\ell_Q \mathcal{E}_Q/\eta$, where $\mathcal{E}_Q$ is the characteristic nematic energy density scale ($\mathcal{E}_Q=$\cut{$0.3$}\add{$0.1$}) and $\ell_Q=\sqrt{K/\mathcal{E}_Q}$ is the equilibrium nematic persistence length\add{, which is comparable to the defect core radius}. 
The competition between these two velocity scales defines a characteristic self-motility number $V=v_{+1/2}/\nu_Q \sim$\cut{$h\zeta/\sqrt{K \mathcal{E}_Q}$}\add{$\ell_Q h\zeta/K$}. 
\add{Considering microtubule-kinesin bundle suspensions, \phalf disclinations move with a self-motility~\cite{Dogic2015} of $v_{+1/2}\approx 8\mu\text{m}\cdot\text{s}^{-1}$.
An order of magnitude estimate of $\nu_Q\sim 10\mu\text{m}\cdot\text{s}^{-1}$ is found by estimating that the nematic elastic constant is $K\sim1\text{pN}$, that the dense suspension at an oil/water interface has a viscosity roughly an order of magnitude larger than water, and that the nematic persistence length is approximately the observed defect core size $\ell_Q\sim10\mu\text{m}$~\cite{Dogic2015}.
Consequently, $V\sim1$ is expected in such systems.} 

By plotting flow structures as a function of activity number $A$ and self-motility number $V$, the flow regimes are well separated into distinct dynamical steady states (\fig{fig:phaseDiagram}). 
The non-flowing quiescent state is shown at the lower left and meso-scale turbulence at the upper right, with unidirectional flow, oscillating flow and Ceilidh dynamics found in between. 
The oscillating-flow state forms one boundary with the Ceilidh dynamics and the transition does not depend strongly on activity number $A$ at moderate $V$ (\fig{fig:phaseDiagram}). 
Likewise, the first measurable occurances of active puffs denoting the early onset of meso-scale from Ceilidh dynamics is seen to occur at an activity number $A$ that depends only weakly on $V$. 
\cut{The locations of the transitions were verified to be independent of both initial conditions and nematic anchoring at the walls.}
 
The region of the dynamical phase diagram (\fig{fig:phaseDiagram}) that we can access in our simulations is limited by numerical stability and it would be interesting to reach further into the low $A$/high $V$ and high $A$/low $V$ regimes. 
In particular the form of the phase boundaries suggest that a direct transition from the oscillating flow state might occur at higher self-motility. 
While we have simulations in the low $V$/high $A$ region of the diagram, it is difficult to clearly identify Ceilidh dynamics prior to the transition to meso-scale turbulence. 
While our previous work has shown that the transition from the Ceilidh dance to meso-scale turbulence belongs to the directed-percolation universality class~\cite{Doostmohammadi2016turb}, this is suggestive of a secondary route to turbulence that goes directly from oscillating flows to chaotic flows. 
Furthermore, it will be interesting for future studies to explore the large $V$ and moderate $A$ region of phase space to see if there exists additional routes directly from oscillating to disordered active flow states. 

\begin{figure}
	\centering
	\includegraphics[width=0.475\textwidth]{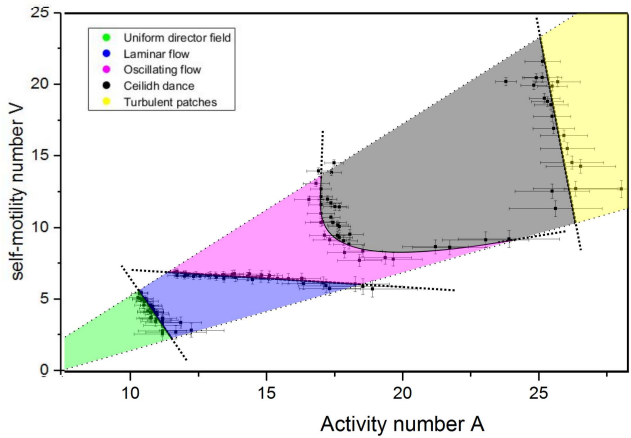}
	\caption{{\bf Dynamical steady state diagram of flow structures.} Flow regimes as a function of the dimensionless activity number $A = h/\ell_\zeta$\cut{$= \sqrt{\zeta h^2/K}$} and self-motility number $V=v_{+1/2}/\nu_{Q}$\cut{$=A\sqrt{\zeta/\mathcal{E}_Q}$}. The transitions are determined by increasing the activity in increments of $\Delta\zeta=2\times10^{-4}$.}
	\label{fig:phaseDiagram}
\end{figure}

\subsection{Geometries that eliminate periodic boundary conditions}

\begin{figure}
  \centering
  \includegraphics[width=0.475\textwidth]{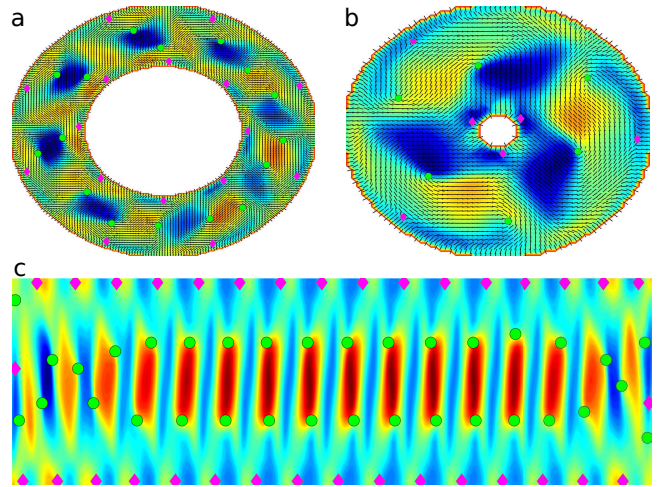}
  \caption{{\bf Ceilidh dynamics in systems that do not use periodic boundary conditions.} 
  \add{(a) Ceilidh dynamics in an annular geometry with an outer radius of $R_o=56$. 
  The difference between the inner and outer radii is $h=25$. 
  (b) The Ceilidh dynamics remain stable even for the large curvature of $R_o=31$ and $h=25$. 
  (c) Ceilidh dynamics in a closed box of height $h=25$ and length $L=500$. Identical to the channel walls, the lateral boundaries are impermeable, no-slip walls with strong homeotropic anchoring.
  }
  }
  \label{fig:geometries}
\end{figure}

\add{The simulations presented thus far are all for a channel geometry with periodic boundary conditions. 
To test whether Ceilidh dynamics arise in experimentally realizable geometries, we consider two bounded geometries: annuli (\fig{fig:geometries}(a-b)) and closed channels (\fig{fig:geometries}(c)). 
For consistency with the previous results using periodic boundary conditions, we keep anchoring at the boundaries, initial conditions and physical parameters the same, and set the gap between parallel confining walls to $h=25$ in all cases.}

\paragraph{Annuli} 
\add{As expected, annular geometries of two concentric bounding circles with small curvature (\fig{fig:geometries}(a)) correspond closely to the results for Ceilidh dynamics in a channel with periodic boundary conditions: 
Non-motile \mhalf disclinations are evenly spaced around the inner and outer circular walls, while self-propelled \phalf disclinations travel azimuthally around the annular gap with half moving clockwise and half counter-clockwise. 
The vortex lattice is preserved~\cite{Theillard2016} and remains relatively unperturbed from the results of the channel geometry with periodic boundary conditions. 
Further increasing the curvature does not cause the Ceilidh dynamics to become unstable, though the vortex lattice skews substantially (\fig{fig:geometries}(b)) and becomes comparable to the vortex fields observed in the vicinity of moderately confined rotors~\cite{Thampi2016}.} 

\paragraph{Closed channels} \add{We next remove the periodicity altogether by considering a channel geometry that is closed by impermeable no-slip lateral walls (\fig{fig:geometries}(c). 
We observe well-ordered Ceilidh dynamics in the centre of the channel, far from the lateral walls. 
Ceilidh dynamics emerge in the center of closed rectangular geometries with a sufficiently large aspect ratios ($\sim10$). 
Near the lateral walls (within $\sim4$ vortex sizes), a disorderly flow pattern occurs. 
In the immediate vicinity of the lateral walls disclination pairs are created, which feed \phalf disclinations into the the ordered Ceilidh dancing in the centre of the channel. 
The resulting \mhalf disclination that arises from such pair-creation events drifts toward the adjacent lateral wall until it is annihilated by an outgoing \phalf disclination. 
}

\subsection{Lattice defects}

\begin{figure*}
	\centering
	\includegraphics[width=0.95\textwidth]{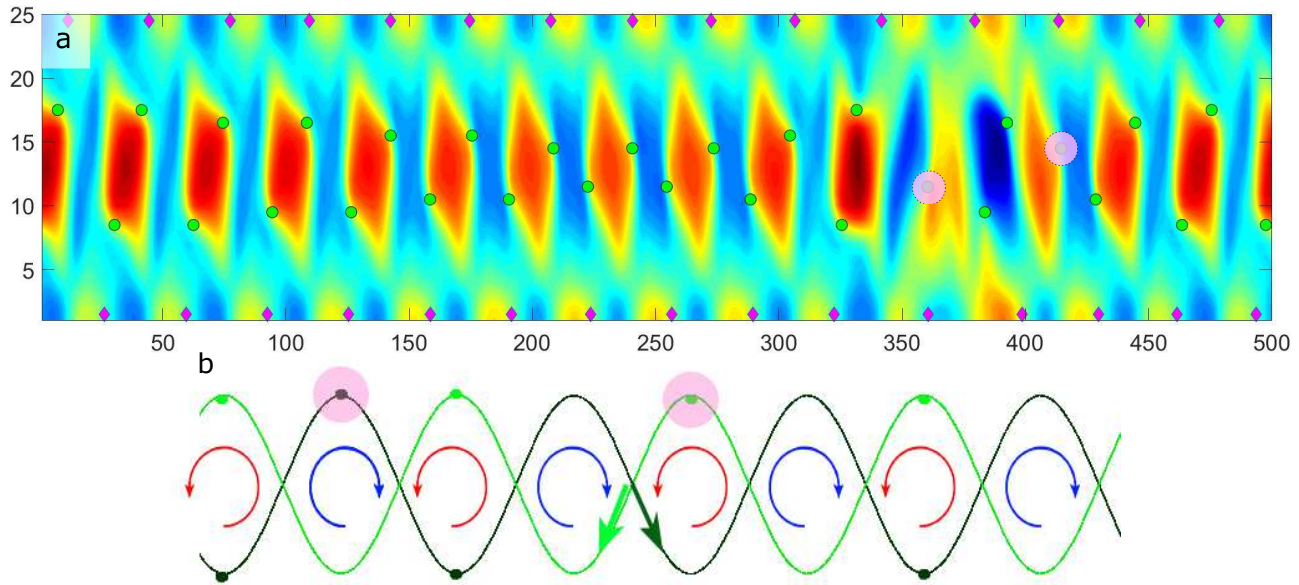}
	\caption{{\bf Broken-pair lattice defect (see \movie{3}).} (a) Simulation snapshot of the instantaneous vorticity field of the ordered vortex lattice, and topological disclinations (\mhalf disclinations as magenta diamonds and \phalf as green dots). A broken-pair lattice defect exists at the ends of the channel (highlighted in magenta). (b) Schematic of the broken-pair lattice defect. 
	The dynamically changing pair of separated \phalf disclinations (highlighted in magenta) are separated by other paired-dancers. The number of disclinations moving leftward (light green dots) remains equal to the number moving rightward (dark green dots).
	}
	\label{fig:brokenpairdefect}
\end{figure*}

As a system-spanning ordered flow state, the Ceilidh dynamics state itself can possess various irregularities. 
We refer to these imperfections in the long-range structure as \emph{lattice defects}, which should not be confused with the \pmhalf topological disclinations that occur in the continuous director field. 
These lattice defects arise when the system becomes trapped in a state in which a dancing pair of \phalf disclinations is separated by other dancing-pairs. 
While  the majority of irregularities are resolved as the steady state is approached, large elastic energy barriers can hinder rejoining of separated pairs and results in long-lived lattice defects. 
\add{In channels with periodic boundary conditions, we} do not observe additional disclinations (the number of topological disclination pairs is still equal to the number of vortices). 
Rather, there are only integer numbers of irregularities in the dynamic dance of the traveling \phalf disclinations.

\subsubsection{Broken-pair lattice defect}

There is a finite probability that the distance between a pair of topological \phalf disclinations does not reduce to $\lesssim h$ before another dancing-pair forms between them. 
The pair is then separated by a large energy barrier and we term this situation a \emph{broken-pair lattice defect} (\fig{fig:brokenpairdefect}; \movie{3}). 
In this broken-pair lattice defect, the number of disclinations moving leftward remains equal to the number moving rightward. 
Due to elastic repulsion between the like-signed disclinations and the confinement of the narrow channel, the separated disclinations cannot escape this imperfect state. 
However, it is important to recognize that no particular dancing-disclination pair permanently makes up the broken-pair lattice defect. 
Rather the Ceilidh dance ensures different dancing disclinations pass through the broken-pair lattice defect. 
In doing this, the disclinations shear the local vortex, causing it to become momentarily skewed and stronger (\fig{fig:brokenpairdefect}; \movie{3}). 
Though the local vorticity field within the broken-pair lattice defect region is perturbed, the globally ordered topological dynamics of the rest of the system is not strongly affected.

\subsubsection{Drift-lattice defect}

In the broken-pair lattice defect, the number of disclinations on the leftward moving path remains equal to the number moving rightward. 
However, it is possible that upon the formation of the vortex lattice more disclinations reach one path than the other (\fig{fig:driftdefect}(a); inset). 
As in the case of the broken-pair lattice defect, the energy cost of escaping this state is high and no disclination is observed to be able to change directions by switching dance-paths. 
In this case, two more disclinations move in one (spontaneously chosen) direction, and their activity drives a net flow $v_\textmd{drift}^{(1)}$ in that direction in addition to the zero-averaged vortex lattice flow. 
For this reason, we term this configuration a \emph{drift-lattice defect} (\movie{4}). 

Although, increasing the activity is not seen to raise the probability of producing a drift-lattice defect in the Ceilidh state (\fig{fig:driftdefect}(b)), increasing the length of the channel makes lattice defects more likely. 
For long enough channels, there is a non-zero probability that the number of lattice defects is greater than one. 
For each additional drift-lattice defect in the system there are two more \phalf disclinations moving in the spontaneously chosen direction.
In this way, the vortex lattice can be seen as a ground state and drift-lattice defects as excited states with quantization number $n$. 
The net flux of the system is seen to be cumulative, such that $v_\textmd{drift}^{(n)} = n v_\textmd{drift}^{(1)}$. 
The maximum number of drift excitations observed in our simulations is $n=3$ (\fig{fig:driftdefect}(a); \movie{4}(b)). 
We \add{would} expect that the probability of obtaining an ideally ordered system will decrease\cut{ and $n \geq 1$}\add{, that $n \geq 1$ will} become more probable\add{, and that a well defined number density of localized lattice defects will emerge} as the length of the channel is increased. 

The drift velocity $v_\textmd{drift}^{(1)}$ due to a single drift-lattice defect can be determined by estimating the net active force and viscous drag. 
The active force per unit volume $-\zeta \bnabla\cdot\tens{Q}$ is most significant near the $\pm 1/2$ disclinations, whose local deformation can be approximated as a solitary defect in bulk, $\vec{n}_{\pm1/2} = \left[ \cos\left(\pm\tfrac{1}{2}\phi\right) , \sin\left(\pm\tfrac{1}{2}\phi\right) \right]$. 
By assuming that the ideal \phalf disclinations are located at and oriented along the centerline of the channel and that the strong anchoring of the walls screen director deformations beyond a distance of $\sim h$, the net active force on the fluid due to a drift-lattice defect is found to be $F^{(1)}_\text{act} = c_a \eta v_{+1/2}$, where the numerical coefficient $c_a$ results from the precise integration area. 
For the case of $n$ drift lattice defects, the net force increases to $F^{(n)}_\text{act} = n \; F^{(1)}_\text{act}$. 
The drag that balances this active force is approximated by realizing that the train of vortices in the mid-channel region move \textit{en masse} (\movie{4}), while the flow in the near-wall region must viscously decrease from $\approx v_\textmd{drift}^{(n)}$ to zero at the no-slip boundary. 
Approximating the flow as simple shear and integrating over the near-wall regions gives the drag force to be $F_\text{drag} =- \left( c_d L/h \right) \eta v_\text{drift}^{(n)}$, where $c_d$ is a numerical drag coefficient whose details depend on the zero-averaged rotating vortex size and structure within the mid-channel region. 
These estimates predict a quantized drift velocity $ v_\text{drift}^{(n)} \sim \left(\tfrac{h}{L}\right) n v_{+1/2}$ in agreement with \fig{fig:driftdefect}(a). 

\section{\cut{Discussion}\add{Conclusions}}

It has been widely recognized that activity can drive collective motion in biological systems and the work presented here stresses the fact that activity can also produce highly ordered, but simultaneously dynamic, flow and topological fields. 
The topological Ceilidh dynamics that we report emerge as an intermediate state between spontaneous unidirectional flow and meso-scale turbulence for active nematics within a confining channel\add{, when the characteristic vorticity length scale of meso-scale turbulence is comparable but smaller than the channel height}. 
\cut{The dancing pair dynamics of \phalf disclinations is accompanied by a well defined vortex lattice
The characteristic length scale of the vortex lattice, which is set by a competition between the activity and the nematic elasticity, must be comparable but smaller than the channel height. When the height is too large, meso-scale turbulence arises within the channel. The characteristic correlation length scales in meso-scale turbulence are well known experimentally for active microtubule bundles~\cite{Dogic2012} and dense bacteria suspensions~\cite{Dunkel2013} to be on the order of $\sim10-100~\mu\textmd{m}$, suggesting that the Ceilidh dynamics reported here may be observable in active topological microfluidics devices.} 

While the time-varying topological disclination dynamics of the Ceilidh dance state were unexpected, \cut{our work is not the first to report an ordered vortex lattice. 
The} \add{the} emergence of vortex lattices in active matter have been observed experimentally in motility assays of microtubles~\cite{Chate2012} and self-propelled particles~\add{\cite{Grossmann2014,Jiang2017}} with short range interactions and also numerically by hydrodynamic screening of activity-induced flows due to frictional damping~\cite{Doostmohammadi2016NatComm}. 
Indeed, repeating vortex patterns have recently been reported in dense suspensions of bacteria confined within microfluidic channels~\cite{Wioland2016}, although the possibility of dynamically ordered disclination states was not discussed. 
In each of these examples, flows are screened.
As such, the Ceilidh dance can be interpreted as arising due to the screening of the flow by the microfluidic channel walls. 

Previous works have reported the emergence of long-range orientational (but not positional) order of active topological disclinations~\cite{Dogic2015}; the current work advances this research direction towards dynamically ordered states and the effect of structural defects in the disclination patterning. Just as impurities, vacancies, dislocations and other defects play an important part in traditional condensed matter systems, the active-fluidic lattice defects discovered here may play critical but previously overlooked roles in precisely controlling the collective dynamics of biological systems, such as cellular monolayers or embryonic epithelial layers and will be relevant to the design of active microfluidic devices. 

\begin{figure}
	\centering
	\includegraphics[width=0.475\textwidth]{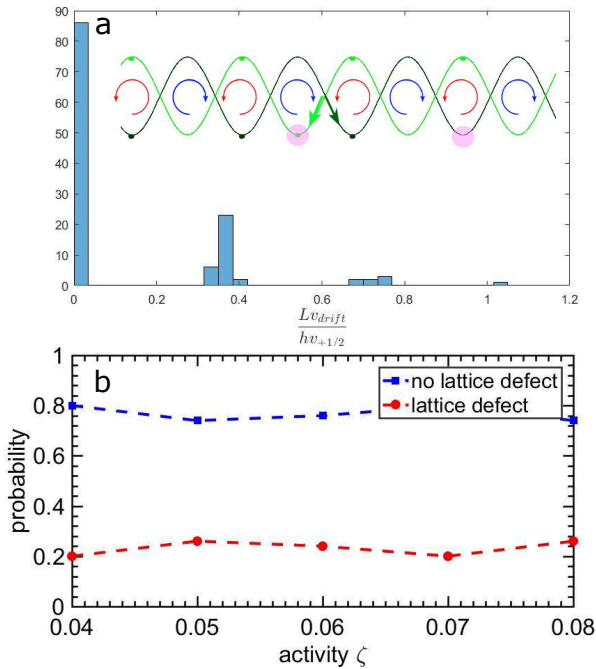}
	\caption{{\bf Quantization of drift velocity due to lattice drift defects (see \movie{4}).} (a) The distribution of net drift velocity is a series of evenly spread peaks corresponding to the number of drift lattice defects present in a given realization. 
	For these finite-sized channels the most probable state is the ideal Ceilidh dance with no lattice defects.
	(inset) Schematic of the drift-lattice defect. 
	An unequal number of \phalf defects travels in a spontaneously determined direction along the left and rightward moving trajectories. 
	This asymmetry drives a net flow $v_\text{drift}$.
	(b) Probability of finding a lattice defect (broken-pair or drift lattice defect) as a function of activity. The likelihood of a simulation containing a lattice defect is not observed to depend on the activity. 
	}
	\label{fig:driftdefect}
\end{figure}

\begin{acknowledgments}
\textbf{Acknowledgments:} This work was supported through funding from the ERC Advanced Grant 291234 MiCE and we acknowledge EMBO funding to TNS (ALTF181-2013). We thank Hugo Wioland for helpful discussions. 
\end{acknowledgments}

\appendix
\section{Movie captions}

\begin{description}
 \item[Movie 1] {\bf Dancing of topological Ceilidh dynamics.} (a) Vorticity field $\vec{\omega}\left(\vec{r},t\right)$ and streamlines of the highly ordered vortex lattice. 
 The dynamically ordered state of disclinations within the channel is shown as \mhalf topological disclinations near the channel walls (magenta diamonds) and \phalf disclinations transversing the mid-region (green dots). 
 The system has zero net-topological charge and equal numbers of \phalf disclinations travel to the left as to the right.
 (b) The scalar order field $q\left(\vec{r},t\right)$ and director field $\vec{n}\left(\vec{r},t\right)$ for the same simulation as \movie{1}(a). 
 \item[\movie{2}] {\bf Vorticity field $\vec{\omega}\left(\vec{r},t\right)$ and and topological defects for the four types of flow states observed in simulations of active nematic fluids confined within two dimensional micro-channels.} 
 Magenta diamonds denote \mhalf topological disclinations and self-motile \phalf disclinations are represented as green dots. 
 (a) Unidirectional flow in a spontaneously chosen direction along the channel occurs at low activity. 
 (b) Oscillating but unidirectional laminar flow, with temporary traveling vorticity patterns that intermittently dissipate then re-form. 
 (c) Ceilidh dynamics with a well-defined vortex lattice at intermediate activity. 
 (d) The chaotic flow state of meso-scale turbulence with the continual creation and annihilation of topological disclinations at the highest activity numbers. 
 \item[\movie{3}] {\bf Broken-pair lattice defect.} Vorticity field $\vec{\omega}\left(\vec{r},t\right)$ and topological disclinations (\mhalf magenta diamonds; \phalf green dots) for topological Ceilidh dynamics with a single broken-pair lattice defect, in which the number of disclinations moving leftward remains equal to the number moving rightward. 
 The lattice defect locally skews the vorticity field about the two separated \phalf disclinations. 
 \item[\movie{4}] {\bf Drift-lattice defects for a single defect (top) and for three defects.} 
 (a) Vorticity field $\vec{\omega}\left(\vec{r},t\right)$ and topological disclinations (\mhalf magenta diamonds; \phalf green dots) for Ceilidh dynamics with a single drift-lattice defect ($n=1$), in which two more disclinations are moving leftward than rightward. 
 A vertical dashed line is drawn across the channel to mark a static point against which the leftward drift of the vortex lattice can be identified. 
 Alternatively, an arrow points to a single \mhalf defect so that the distance travelled can be assessed at all times. 
 (b) The same system as \movie{4}(a) but for an example simulation with three drift-lattice defects ($n=3$). Comparison with \movie{4}(a) shows that the drift velocity is substantially faster than when only a single drift-lattice defect is present. 
\end{description}

\bibliography{ceilidh}

\end{document}